\documentclass[prl,aps,preprint]{revtex4-1}%

\usepackage{amsmath}
\usepackage{graphicx} 
\usepackage{bm} 
\usepackage{color} 
\usepackage{amssymb}
\usepackage{verbatim}
\usepackage[colorlinks=true, pdfstartview=FitV, linkcolor=blue, citecolor=blue, urlcolor=blue]{hyperref} 

\newcommand{\ee}[1]{\times 10^{#1}}
\newcommand{\mr}[1]{\mathrm{#1}}
\newcommand{\unit}[1]{\,\mathrm{#1}}
\newcommand{\um}{\,\mu{\rm m}}

\newcommand{\muB}{\mu_{\rm B}}

\newcommand{\fc}{f_c}
\newcommand{\kc}{k_c}
\newcommand{\Tc}{T_c}
\newcommand{\Tramp}{T_\mr{ramp}}
\newcommand{\vecB}{\bm{\mr{B}}}

\newcommand{\vecM}{\bm{\mr{M}}}
\newcommand{\vecF}{\bm{\mr{F}}}
\newcommand{\vecmu}{\bm{\mr{\mu}}}
\newcommand{\vecr}{\bm{\mr{r}}}

\newcommand{\Fmin}{F_\mr{min}}
\newcommand{\dx}{\partial_x}

\begin{document}

\title{Ultrasensitive mechanical detection of magnetic moment using a commercial disk drive write head}
\author{Y. Tao}
\author{A. Eichler}
\author{T. Holzherr}
\author{C. L. Degen}
\email{degenc@ethz.ch}
%
\affiliation{Department of Physics, ETH Zurich, Otto Stern Weg 1, 8093 Zurich, Switzerland}
\date{\today}
\begin{abstract}
Sensitive detection of weak magnetic moments is an essential capability in many areas of nanoscale science and technology, including nanomagnetism, quantum readout of spins, and nanoscale magnetic resonance imaging.
Here, we show that the write head of a commercial hard drive may enable significant advances in nanoscale spin detection.  By approaching a sharp diamond tip to within $5$\,nm from the pole and measuring the induced diamagnetic moment with a nanomechanical force transducer, we demonstrate a spin sensitivity of $0.032$ Bohr magnetons per root Hz, equivalent to $21$ proton magnetic moments.  The high sensitivity is enabled in part by the pole's strong magnetic gradient of up to $28\times 10^6\unit{T m^{-1}}$ and in part by the absence of non-contact friction due to the extremely flat writer surface.  In addition, we demonstrate quantitative imaging of the pole field with $\sim 10\unit{nm}$ spatial resolution. We foresee diverse applications for write heads in experimental condensed matter physics, especially in spintronics, ultrafast spin manipulation, and mesoscopic physics.
\end{abstract}

\maketitle

\section{Introduction}

Magnetic recording heads generate intense local magnetic field pulses to write bits of information onto a magnetic medium.  Bits are encoded in the magnetization direction of magnetic domains separated by less than $20\unit{nm}$ in state-of-the-art devices.  Write fields thus must be confined to a very narrow region in space, leading to extremely high local gradients.  Although these gradients cannot be precisely measured, they are estimated to exceed $20\ee{6}\unit{T m^{-1}}$ \cite{tsang06}.  This is far beyond the capability of static magnetic tips ($\sim 5\ee{6}\unit{T m^{-1}}$ \cite{poggio07,mamin12apl}) and that of nanoscale coils or microstrips ($\sim 10^5\unit{T m^{-1}}$ \cite{poggio07,nichol13}), and likely close to the highest experimentally achievable gradient for this kind of field source.  In addition, write poles are rapidly switchable, potentially allowing for dynamical control of magnetic fields up to $1\unit{GHz}$ \cite{george03,xing07} at low power consumption.  With these features, the hard drive industry has created a tool that could enable important advances in many areas of nanoscale experimental physics.

A significant challenge in characterizing and exploiting writer fields is their nanometer spatial confinement.  In this study, we introduce a new variant of force microscopy -- magnetic susceptibility force microscopy ($\chi$FM) -- to localize and measure the write pole field quantitatively and with high spatial resolution.  The technique relies on the small diamagnetic moment induced in a nominally non-magnetic diamond tip \cite{tao15nl} that is positioned over the pole.  We show that by using a state-of-the-art nanomechanical force transducer \cite{chui03,tao15nanotech}, tip magnetic moments far below one Bohr magneton can be detected, thus greatly advancing the sensitivity of mechanical spin detection \cite{rugar04,mamin09,poggio10}.  Moreover, because of the diamond tip's well-defined material composition and extremely sharp end radius ($<10\unit{nm}$), quantitative and high-resolution field maps of the pole can be reconstructed.  Our method thus provides advantages over magnetic force microscopy (MFM) \cite{koblischka09,amos10,lu10,li12,tanaka12} and electron holography \cite{einsle15}, which are difficult to quantify, barely reach sufficient resolution, or provide two-dimensional projections.

\section{Results}

\subsection{Experimental Setup and Measurement Technique}

\begin{figure}[h!]
\includegraphics[width=\columnwidth]{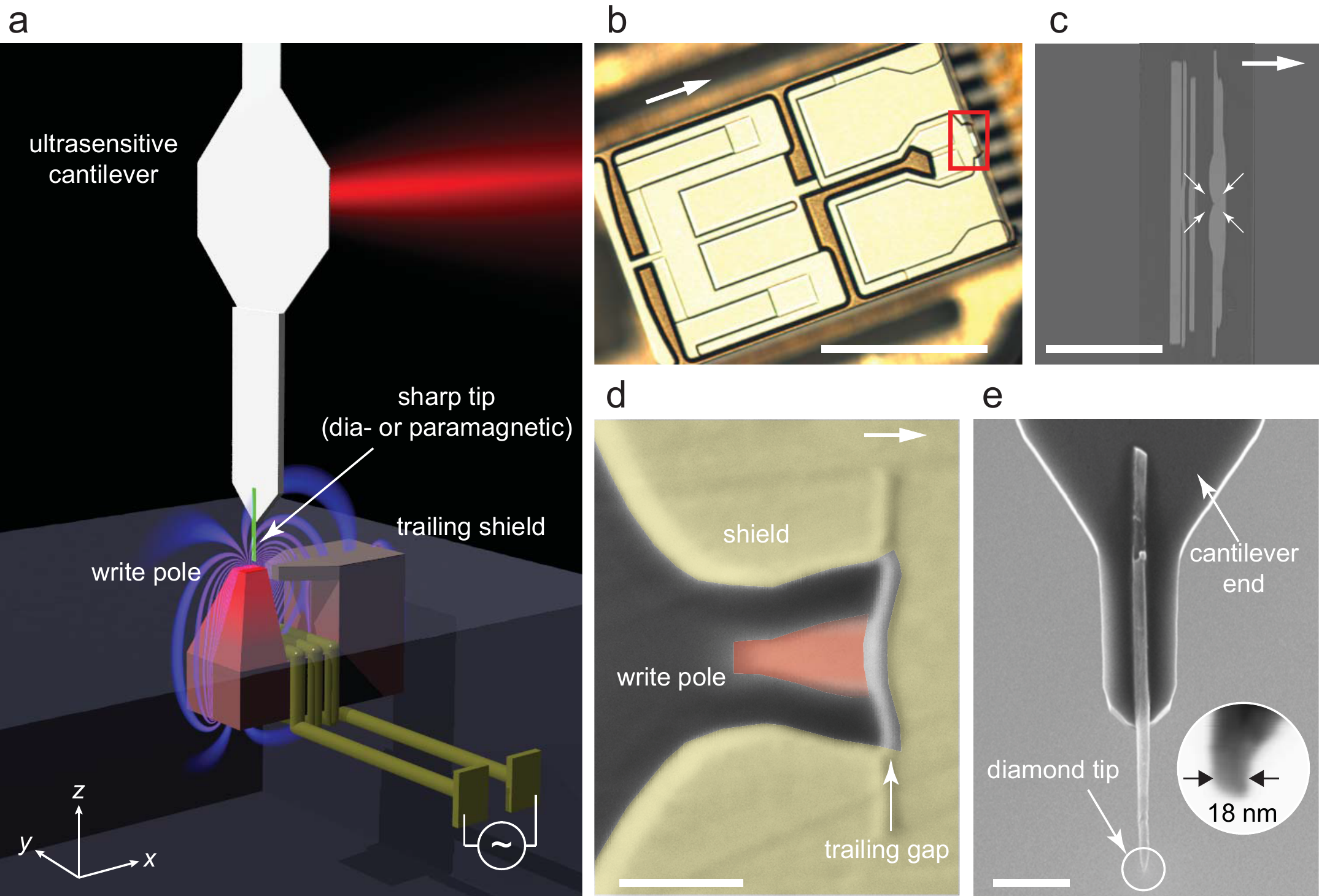}
\caption{
{\bf Geometry of write head experiment and device.}
{\bf a}, A sharp diamond needle (green), attached to a nanomechanical force transducer, is positioned over the write pole of a magnetic recording head.  An alternating current periodically switches the pole polarity and induces magnetic gradient forces through dia- or paramagnetism in the tip.  Experiments are carried out in a custom scanning force microscope operating at 4 Kelvin and in high vacuum.
{\bf b}, Optical micrograph of the write head device.  Arrows in b, c and d point in the direction of the trailing edge (in positive $x$ direction). The write head was extracted from a commercial Seagate hard drive, reconnected to external leads and mounted in the apparatus as discussed in the Methods section. Scale bar is $0.5$\,mm.
{\bf c}, Zoom-in on the write/read region of the device. The write pole is at the center of the four arrows. Scale bar is $20\,\mu$m.
{\bf d}, The $\sim 90\times 60\unit{nm^2}$ write pole (red) is surrounded by a return shield (yellow) that serves to recollect the field lines. The gradient is largest in the $\sim 20\unit{nm}$-wide trailing gap between pole and shield. Scale bar is $100$\,nm.
{\bf e}, Diamond nanowire probe attached to the end of an audio-frequency ($\sim 5\unit{kHz}$) silicon cantilever.  Inset shows apex of tip B. Scale bar is $2\,\mu$m.
}
\label{fig:Fig_1}
\end{figure}

\begin{figure}[h!]
\includegraphics[width=0.4\columnwidth]{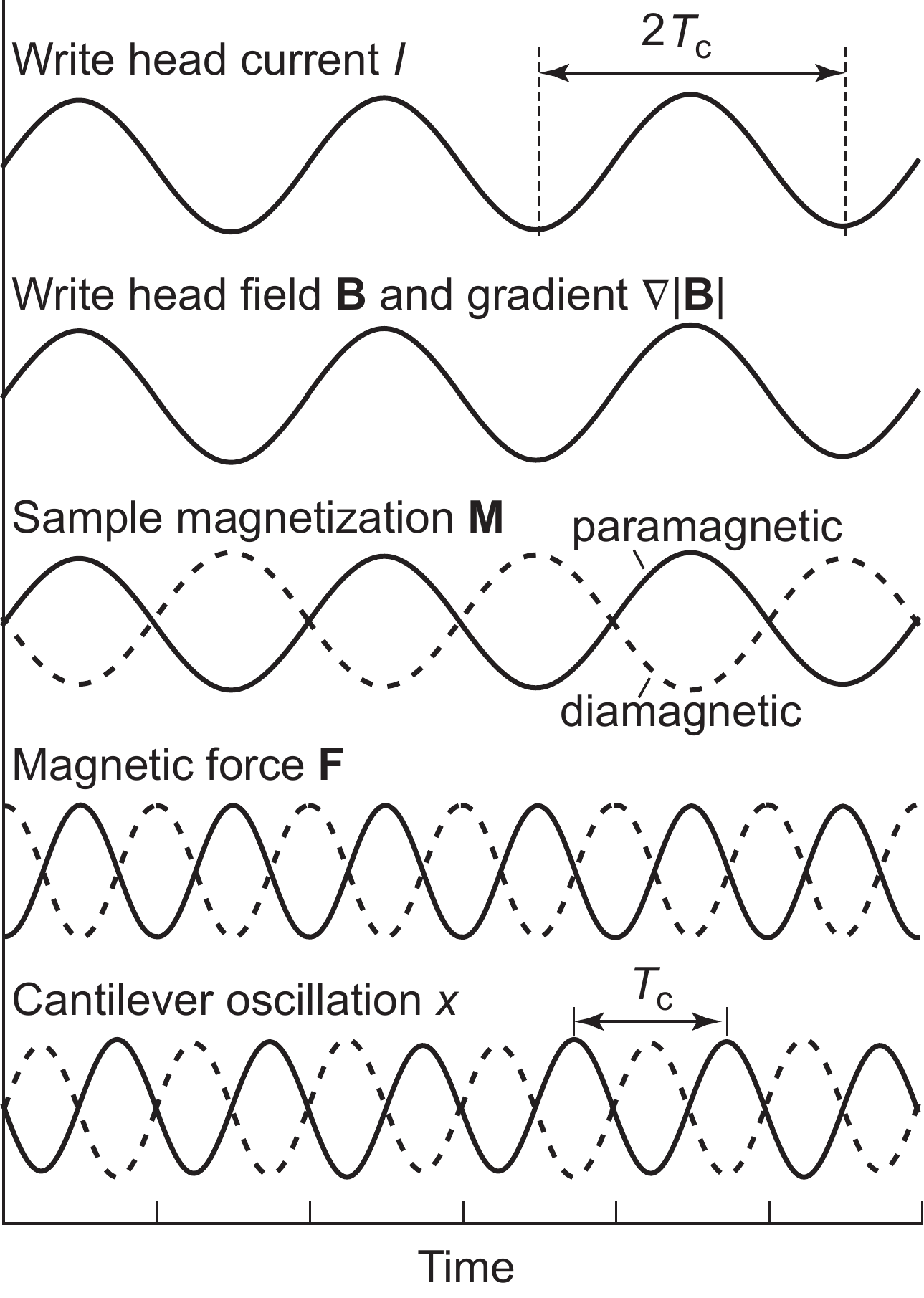}
\caption{
{\bf Basic driving protocol for magnetic susceptibility force microscopy ($\chi$FM).}
A sinusoidal current with frequency $\fc/2$ is applied to the coil of the write element, generating an oscillating pole magnetic field $\vecB$ and gradient $\nabla|\vecB|$. The oscillating field in turn induces an oscillating sample magnetization $\vecM=\chi \vecB/\mu_0$. The sample with volume $V$ experiences a magnetic force $\vecF=V \nabla(\vecM\cdot\vecB)$ that drives the mechanical resonance $\fc$ of an ultrasensitive cantilever force transducer. The amplitude of the resulting cantilever oscillation $x$ is proportional to the force. $T_c=1/\fc$ is one cantilever oscillation period.
}
\label{fig:Fig_2}
\end{figure}

The experimental geometry and the basic protocol for magnetic susceptibility force measurements are presented in figures \ref{fig:Fig_1} and \ref{fig:Fig_2}, respectively.  The magnetic force is generated by passing an alternating current $I$ through the write head's drive coil.  The current dynamically magnetizes the write pole and a stray field $\vecB$ appears above the pole.  A tip placed in the stray field acquires a small magnetization $\vecM=\chi \vecB/\mu_0$ owing to its magnetic susceptibility $\chi$ ($\mu_0 = 4\pi \times 10^{-7}$\,NA$^{-2}$ is the permeability of free space and typically $|\chi|<10^{-4}$).  A weak force develops that attracts or repels the tip from the region of strongest field, depending on whether the tip is paramagnetic ($\chi>0$) or diamagnetic ($\chi<0$).  For a point-like particle located at position $\vecr$ the force is
\begin{equation}
\vecF(\vecr) = \nabla[\vecmu(\vecr)\cdot \vecB(\vecr)] = \frac{\chi V}{\mu_0} \nabla |\vecB(\vecr)|^2 \ ,
\label{eq:force}
\end{equation}
where $\vecmu = V \vecM = V \chi \vecB/\mu_0$ is the magnetic moment and $V$ the volume of the particle.  For a transducer responsive to forces along the $x$-direction, the measured force signal is given by $F_x(\vecr) = \chi V \partial_x |\vecB(\vecr)|^2/\mu_0 = 2\chi V |\vecB(\vecr)| \dx|\vecB(\vecr)| /\mu_0$, where $\dx|\vecB| \equiv \partial|\vecB|/\partial x$ is the magnetic gradient in $x$-direction.

Since both the susceptibility $\chi$ and the volume $V\sim (10\unit{nm})^3$ are small, the expected magnetic moment is minute, on the order of one Bohr magneton for a field of a few hundred mT.  Even in a high gradient $>10^{6}\unit{T m^{-1}}$ the diamagnetic force will therefore be less than $10^{-16}\unit{N}$.  To distinguish this force signal from background fluctuations and spurious electrostatic driving, we modulate the drive current at half the cantilever resonance ($\fc/2$) while measuring the force generated at $\fc$ (see Fig. \ref{fig:Fig_2}).  For small driving currents $I$, the write pole is not saturated and the field and gradient are both proportional to $I$.  For a sinusoidal modulation $I(t) = I_0 \sin(\pi\fc t)$ with amplitude $I_0$ the resulting force has frequency components at d.c. and $\fc$,
\begin{equation}
F(t) = -\frac{\chi V}{2\mu_0} \partial_x|B_0|^2 [ 1 + \cos(2\pi\fc t) ],
\end{equation}
where $B_0$ is the amplitude of the stray field.  The force then drives mechanical oscillations $x(t)$ of the cantilever that are detected by optical interferometry (see Methods).  Note that as the current is increased, the write pole eventually becomes saturated resulting in a square-wave response to a sinusoidal drive.  This leads to the appearance of higher order terms in the Fourier series.

\subsection{Scanned Measurements}

\begin{figure}[h!]
\includegraphics[width=0.65\columnwidth]{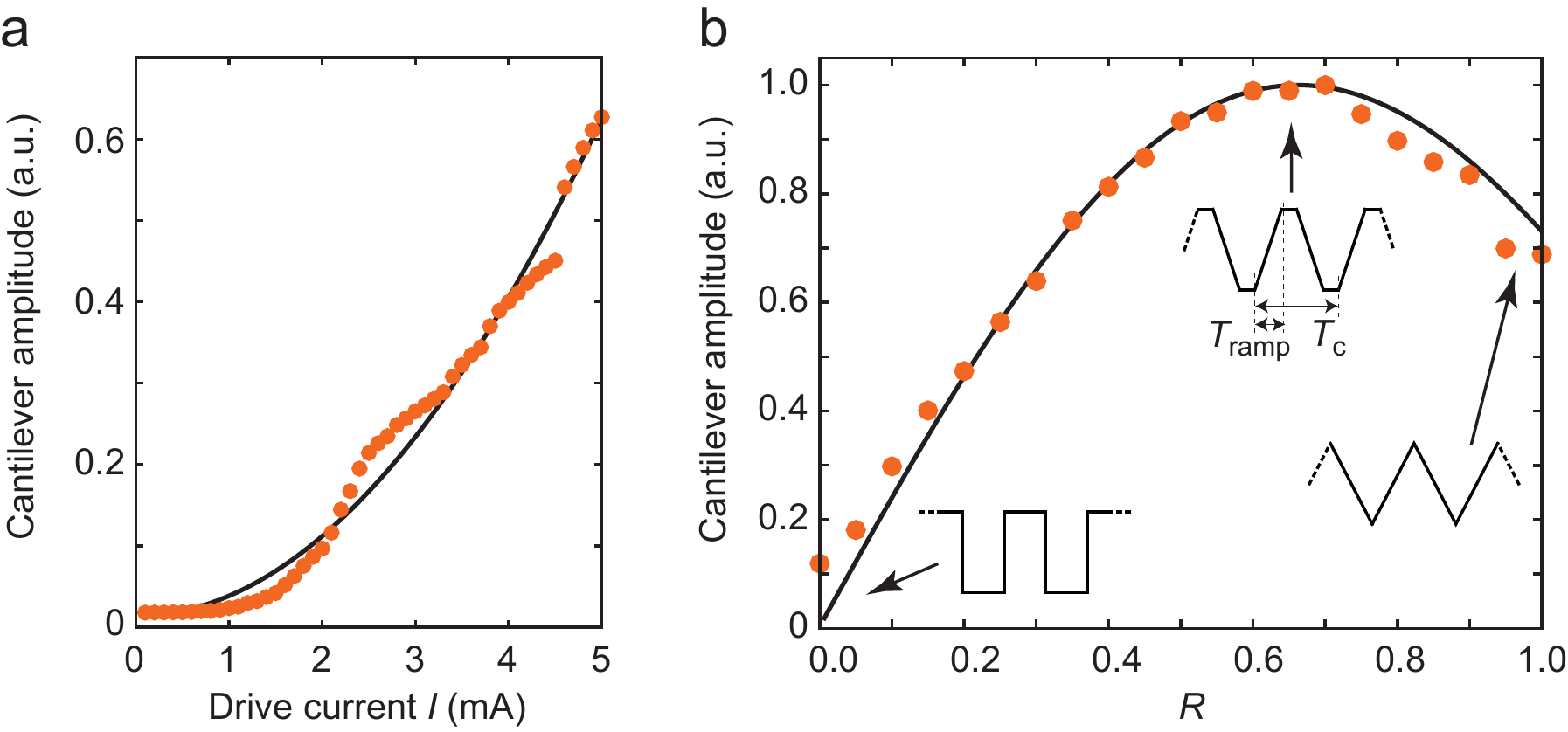}
\caption{
{\bf Demonstration of $\chi$FM.}
{\bf a}, Signal as a function of drive current amplitude, well below magnetic saturation of the pole.  Solid line is $\propto I^2$.  Small deviations from the square law are probably due to domain switching and hysteresis.
{\bf b}, Signal amplitude while changing the current modulation from rectangular ($R=0$) to triangular ($R=1$), as sketched in the figure. $R = 2\Tramp/\Tc$ is the fraction of time spent on ramping the current. 
}
\label{fig:Fig_3}
\end{figure}

In a first experiment, we demonstrate that magnetic susceptibility forces can indeed be measured.  For this purpose, we positioned the diamond tip at $z=20\unit{nm}$ over the pole surface and measured the cantilever signal as a function of drive current amplitude $I$, as shown in Fig. \ref{fig:Fig_3}a.  As expected from Eq. (\ref{eq:force}), the cantilever signal increased quadratically when increasing the driving current from zero to a few mA.  To confirm that the observed signal is truly due to magnetic driving by the write pole, we have also varied the shape of the current modulation.  When using a rectangular drive, no signal was observed, consistent with the absence of a Fourier component for this modulation pattern (Fig. \ref{fig:Fig_3}b).  As the pattern was continuously changed from rectangular to triangular, the cantilever signal first increased, reached a maximum at trapezoidal drive, and then decreased again (see Figure insets).

To record a two-dimensional image, we performed a raster scan over the pole and plotted the cantilever oscillation amplitude as a function of $xy$ tip position (see Fig. \ref{fig:Fig_4}a).  The scan clearly shows two regions of high signal, which we identify as the front and back end of the write pole.  As expected, the largest force is generated near the trailing gap where the bending of field lines is highest.  A finite element simulation of the expected force signal (Fig. \ref{fig:Fig_4}b), based exclusively on the micrograph of Fig. \ref{fig:Fig_1}d, agrees surprisingly well with the scanning experiment (see Supplementary Figure 1 for details).  The signal peaks are localized to within $\sim 10\unit{nm}$, demonstrating the high spatial resolution of the magnetic susceptibility force imaging technique.

By measuring the phase shift between the cantilever oscillation and the drive current, information about the sign of $\chi$ can be extracted in addition to the magnitude of the magnetic force.  Fig. \ref{fig:Fig_4}c,d show two dimensional force images obtained by scanning two types of tips. Tip A was coated with a $10\unit{nm}$ Pt layer and was moderately paramagnetic ($\chi_\mr{Pt}=+2.9\ee{-4}$) \cite{hoare52}.  Tip B was a bare diamond nanowire and weakly diamagnetic ($\chi_\mr{diamond}=-2.2\ee{-5}$) \cite{heremans94}.  The two figures clearly show how the paramagnetic tip is attracted to the center of the write pole (where the magnetic field is highest), while the diamagnetic tip is repelled from the high field region.  These measurements demonstrate that the technique is sensitive to the sign of $\chi$.

\subsection{Reconstruction of Magnetic Field and Gradient}

Although our measurements record images of the magnetic force $F_x$, and not the field $\vecB$, we can rigorously reconstruct the magnetic field and the gradient from a force map.  By integrating Eq. (\ref{eq:force}) along the $x$-direction, one obtains an expression for the magnetostatic energy of the tip as a function of position,
\begin{equation}
W(x) = \int_{-\infty}^{x} dx'F_x(x') = \frac{\chi V}{\mu_0} |\vecB|^2.
\label{eq:energy}
\end{equation}
This expression can be used to deduce the magnetic field, $|\vecB| = [\mu_0 W/(\chi V)]^{1/2}$.  We have quantitatively determined $|\vecB|$ and $\dx|\vecB|$ for the force map of Fig. \ref{fig:Fig_4}d by calibrating the magnitude of the force and the effective tip volume ($V_\mr{eff}=2.9\ee{-24}\unit{m^3}$, tip B) (see Supplementary Figures 2-4 and Supplementary Notes 1-2).  The resulting field maps are shown in Fig. \ref{fig:Fig_5}a,b.  A region of high field is seen over the pole surface, as expected, with a sharp drop of high gradient just to the right.  Despite the relatively low drive current ($I=5\unit{mA}$) and large tip standoff ($z=30\unit{nm}$), magnetic fields and gradients are already quite sizable with peak values around $100\unit{mT}$ and $\sim 3.2\unit{MT m^{-1}}$, respectively.

\subsection{Maximum Achievable Field and Gradient}

To estimate the maximum field and gradient that can possibly be generated at the tip location, we measured the dipole force as a function of tip-surface distance $z$ and as a function of the drive current $I$.  These measurements are also presented in Fig. \ref{fig:Fig_5}.  As the distance between tip and pole surface was reduced, the signal increased until non-contact friction \cite{tao15nl} began to dampen the mechanical oscillation.  The signal decayed exponentially with distance with characteristic length $\delta=11.3\pm0.6\unit{nm}$ for tip B (Fig. \ref{fig:Fig_5}c) and $\delta=13.5\pm1.2\unit{nm}$ for tip A (see Supplementary Figure 5).  The signal decay is thus not influenced by the tip shape.  We found using numerical modeling that $\delta$ is mainly set by the width of the write pole near the trailing gap, which was about $60\unit{nm}$ in our devices.

To test the maximum drive current, we have monitored the signal from tip B while increasing the drive beyond the failure current density (Fig. \ref{fig:Fig_5}d).  The signal continued to increase up to a breakdown current of $\sim30\unit{mA}$, showing that the write pole had not reached magnetic saturation.  We have modeled the magnetic response based on a Langevin magnetization curve with a saturation field of $2.4\unit{T}$ (FeCo) \cite{bardos69} and found that the pole magnetization at $30\unit{mA}$ is about $1.57\pm 0.08\unit{T}$, or roughly 65\% of the saturation magnetization (see Supplementary Note 3).  Note that the power dissipation is small even at the maximum driving amplitude.  For a write head resistance of $R=3\unit{\Omega}$ (that should ideally be achievable), the dissipated power is only $14\unit{mW}$ even when the coil is continuously driven with $30\unit{mA}$.  In our experiments the total dissipated power was in fact limited by lead and contact resistances, and not the write element.  In pulsed mode and at low duty cycle, we expect that saturation can be reached with less than $1\unit{mW}$ average dissipation, eventually permitting Millikelvin operation.

\section{Discussion}

\begin{figure}[h!]
\includegraphics[width=0.70\columnwidth]{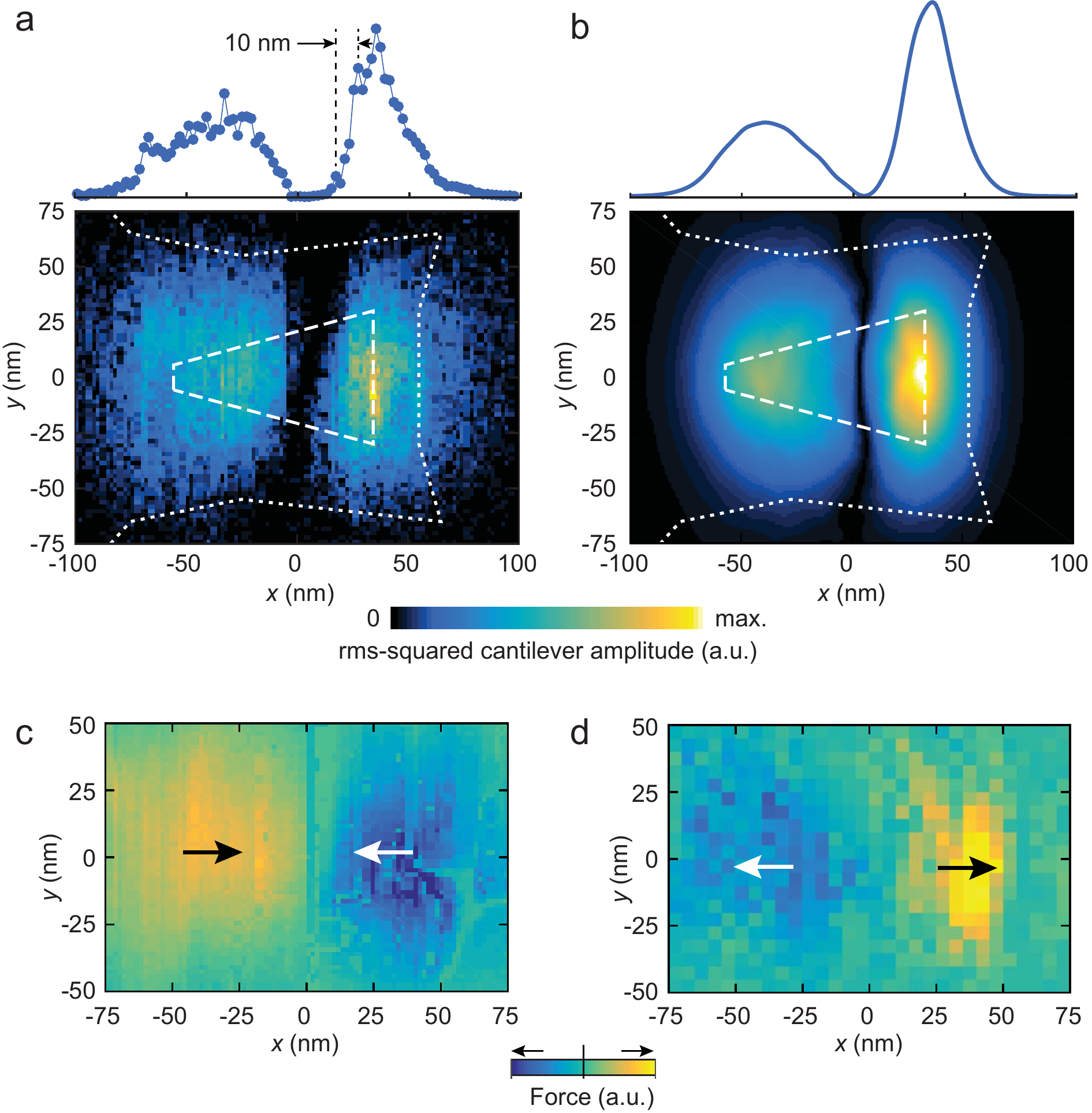}
\caption{
{\bf Two-dimensional imaging of pole field.}
{\bf a}, Experimental force image recorded with tip A.
Two lobes of increased signal are visible at the front and back end of the pole where the gradient $\dx|\vecB|$ is largest. The contours indicate the write pole (dashed) and shield (dotted).  The traces above the image shows the signal intensity (rms-squared cantilever oscillation) along the $x$-direction.
Point spacing is $2\unit{nm}$ with an integration time of $3$ seconds per point.  Drive current was $5\unit{mA}$.
{\bf b}, Independent finite element simulation based on the micrograph from Fig. \ref{fig:Fig_1}d.
{\bf c,d}, Phase-sensitive force images recorded by a paramagnetic Pt-coated tip A (c) and by the diamagnetic bare diamond tip B (d), demonstrating that the sign of $\chi$ can be detected.  Arrows point in the direction of the force.  Color bar scale is linear.  Scan height was $15\unit{nm}$ for tip A and $30\unit{nm}$ for tip B, and drive was $5\unit{mA}$.
}
\label{fig:Fig_4}
\end{figure}

Table \ref{table:gradient} collects some key experimental values for the magnetic field $|\vecB|$ and field gradient $\dx|\vecB|$, representative for a tip standoff of $5\unit{nm}$.  The values are based on the force map in Fig. \ref{fig:Fig_4}b and on the scaling of the force with drive current $I$ and distance $z$.  The experiment demonstrates that magnetic fields around $0.87\unit{T}$ and gradients around $28\unit{MTm^{-1}}$ are present at $30\unit{mA}$ drive, about $5\times$ larger than the gradients observed with static nanoscale ferromagnets \cite{mamin12apl}.  Even larger gradients are expected under magnetic saturation, which could be reached by pulsing or by an external bias field.  The experimental results are consistent with an independent finite element calculation (that shared no common input parameters) which predicts slightly smaller, but overall similar values (see Table \ref{table:gradient}).  The large gradient is a result of the close access ($\sim 5\unit{nm}$) enabled by the diamond nanowire tips, and also thanks to the recessed nature of the pole that prevents oxidation and allows for sharp, highly magnetized pole edges.

\begin{figure}[h!]
\includegraphics[width=0.95\columnwidth]{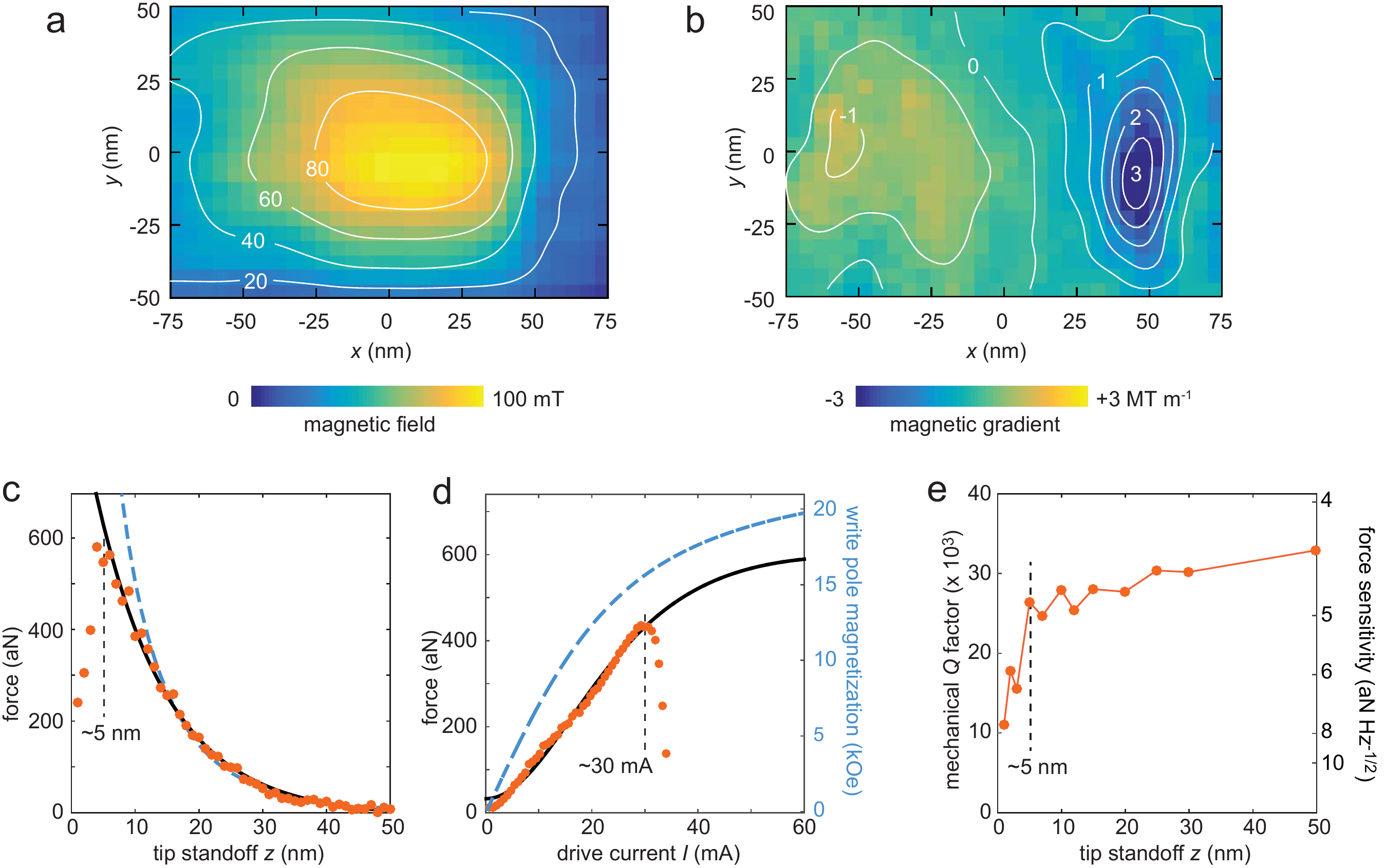}
\caption{
{\bf Quantitative field maps and estimated limits.}
{\bf a,b}, Two-dimensional images of the magnetic field (a) and field gradient (b) for $5\unit{mA}$ drive and $30\unit{nm}$ standoff.  Contour labels are mT and MTm$^{-1}$, respectively.
{\bf c}, Force signal as a function of tip standoff $z$, measured at the $xy$ location with the highest signal.  The solid line is an exponential fit with decay length $\delta = 11.3\pm0.6\unit{nm}$.  The dashed line is the prediction by the finite element model that was vertically scaled to fit the data.  Drive was $12\unit{mA}$.
{\bf d}, Cantilever signal as the drive was increased beyond the breakdown current of $\sim 30\unit{mA}$.  The solid line is a fit based on the magnetization curve of FeCo (see supplementary information).  The dashed line (right scale) is the associated pole magnetization.  Tip standoff was $20\unit{nm}$.  All data are from tip B.
{\bf e}, Mechanical quality factor $Q$ and force sensitivity $\Fmin$ of the transducer as a function of tip standoff $z$.  A force sensitivity of $\sim 5\unit{aN}$ per root Hz is maintained down to $z = 5\unit{nm}$.
}
\label{fig:Fig_5}
\end{figure}

Write head gradient sources could therefore enable important steps forward in mechanical detection of electronic and nuclear spins. In Fig. \ref{fig:Fig_5}c, forces up to $F=568\unit{aN}$ are generated with a net magnetic moment of only $\vecmu = \chi V \vecB/\mu_0 \sim 3.3\ee{-23}\unit{A m}$, roughly equivalent to $\sim 3.6\muB$.  This represents a strong force per electron spin of about $\sim 160\unit{aN/\muB}$.  Since the force sensitivity of the transducer of $\Fmin \sim 5\unit{aN}$ per root Hz is maintained even at $5\unit{nm}$ spacing (see Fig. \ref{fig:Fig_5}e), an excellent magnetic moment sensitivity of $0.032\unit{\muB}$ per root Hz results, which is equivalent to about 21 proton moments per root Hz.  This value improves to $\sim 12$ proton moments if the pole is magnetically saturated.
By comparison, the best reported sensitivities for previous force detectors are $\sim 100$ proton moments \cite{mamin09, degen09}, and  $\sim 250$ proton moments for scanning SQUID sensors \cite{vasyukov13}.  

The sensitivity is even more remarkable since the performance of the force transducer was not particularly optimized in our study.  It could be further enhanced by surface passivation \cite{tao15nanotech,tao16acs}, change of material \cite{teufel09,tao14,moser13}, or lower operating temperatures \cite{mamin01,tao14}.  It is therefore conceivable that write head gradients will pave the way for single nuclear spin detection.  This will constitute a milestone advance towards the realization of the long-standing proposal of single nucleon magnetic resonance imaging \cite{sidles91}.  The depth resolution of such imaging would be limited by the decay length of the pole field, which is $2\delta \sim 23\unit{nm}$ for the write heads used in our study.  This is a significant depth resolution compared to other candidate imaging techniques, like nitrogen-vacancy spin sensors, where single-spin sensitivity does not extend beyond a few nm \cite{staudacher13,mamin13}.

Aside from the detection of magnetic forces, the use of write heads may allow for significant advances in several fields of active research.  Perhaps the most important of these is the manipulation of spin systems in the context of quantum technologies \cite{awschalom13}.  Static magnetic field gradients can be used for rapid spin manipulation by means of electrical fields \cite{tokura06, kawakami14} or quantum dot exchange interaction \cite{coish07, wu04}, and for strong and tunable coupling in hybrid quantum systems \cite{arcizet11}.  The gradient can further be exploited to set up spin registers in quantum simulators \cite{cai13}.  The combination of large field and rapid switching \cite{george03,xing07}, which is difficult to achieve in typical research devices, will allow the implementation of very fast spin manipulation through magnetic resonance techniques \cite{jakobi16}.  At a pulse amplitude of $1\unit{T}$, the nominal spin flip duration is $ 0.5 \times \gamma B_0 / 2\pi \cong 18\unit{ps}$ for electrons and $\sim 12\unit{ns}$ for protons, which would be faster than any previously demonstrated values \cite{yamauchi04,hagaman08,fuchs09}.  These flip rates will eventually be limited by the response time of the write head and pole material \cite{george03,xing07}.

\begin{figure}[h!]
\includegraphics[width=0.65\columnwidth]{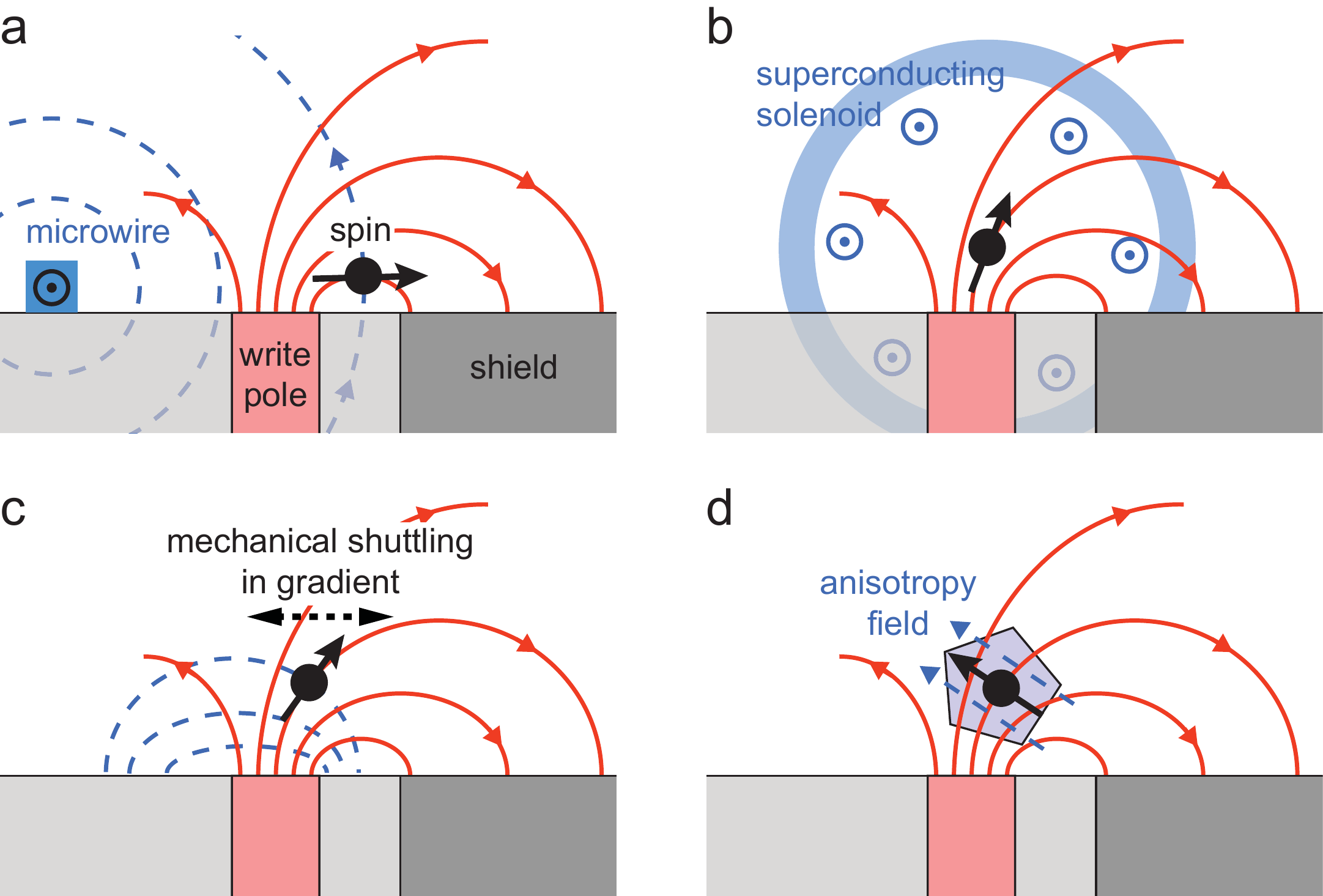}
\caption{
{\bf Proposed arrangements for ultrafast spin manipulation.}
Write head field is symbolized by solid red field lines and orthogonal auxiliary field is symbolized by dashed blue field lines.
{\bf a}, Auxiliary microwire serving as electromagnet \cite{poggio07,nichol13}.
{\bf b}, Auxiliary macroscopic solenoid.
{\bf c}, An ac field is generated by rapid motion of spin in the pole gradient, for example, with a mechanical resonator \cite{arcizet11} or by an electrical field \cite{tokura06, kawakami14}.
{\bf d}, Internal anisotropy provides the orthogonal field, such as the crystal field of a nitrogen-vacancy center \cite{doherty13}.
}
\label{fig:Fig_6}
\end{figure}

Universal spin control requires two orthogonal axes of rotation. Although the write head only provides one axis given by the direction of $\vecB(\vecr)$, there are several possible ways to add a second axis, as proposed in Fig. \ref{fig:Fig_6}.
Beyond spin control, write head devices may finally present creative new opportunities in nanoscale transport.  For example, pulsed spin polarized currents could be launched through local magnetization of ferromagnetic electrodes \cite{wolf01}.  Electrons in confined geometries, such as quantum point contacts, could be locally deflected and the magnetic potential varied within the ballistic regime.  Since write heads have an extremely flat surface made from nonconducting diamond-like carbon, complex lithographic structures including local gates, microwires, constrictions, quantum dots or spin qubits could be conveniently integrated.


\bibliographystyle{naturemag}



\newpage

{\bf Methods:}

{\sl Write heads:}
The hard disc write heads (WH) used in this study were extracted from Seagate Barracuda 1TB and 4TB desktop drives.  As a general rule, we worked on anti-static mats and grounded ourselves while working with the WH to avoid electrostatic discharges.  Care was taken to avoid physical contact with the trailing edge of the WH where the write/read regions are located.

We opened the hard drive casing to gain access to the platter discs and the actuator arms. The WH blocks are located at the end of these arms.  We severed the ends of the actuator arms using a pair of scissors, then peeled the WH block from its metal support with a pair of sharp tweezers. The backside of a WH block has a small circular dab of glue to fix it to the actuator arm. The glue turned out to be problematic for precise, flat positioning of the WH in our experiment.  To remove this dab of dried glue, we flipped the WH block upside down on a clean silicon wafer and scratched the backside of the WH block with a sharp pair of tweezers wrapped in a couple layers of Kimwipe until the back surface was completely clean and residue-free as determined by optical microscopy.  A $\sim 2\unit{nm}$ Pt layer was evaporated on the WH surface to screen electric charges.

For low-temperature scanning probe experiments, we fixed the WH on top of a sample stage made from shapal, an aluminum nitride ceramics with excellent thermal conductivity (see Supplementary Figure 6). We avoided metal as material for the stage to prevent eddy-current damping of radio-frequency magnetic fields that would eventually be used in nuclear spin-manipulation experiments. 

A total of nine metal bonding pads were found on the side of the WH block.  Two of the nine pads were connected to the current-carrying coil inside the WH block for controlling write pole magnetization direction. We identified the two relevant coil leads through trial and error by using magnetic force microscopy for readout verification (see Supplementary Figure 7). For electrical contact to macroscopic wires, we used a silicon jumper chip with lithographically patterned gold thin film structure matching the pad dimensions.  

The mechanical and electric reassembly process took place in the following sequence.  We first glued the jumper chip to the shapal stage using an insulating glue compatible with high vacuum and cryogenic temperatures (Epotek H70E).  We glued copper wires to the two metal pads on the jumper chip using a conducting epoxy (Epotek H20E) and mechanically fixed them with an additional layer of Epotek H70E. Next, the WH block was carefully positioned such that the two leads for the write pole were aligned with the corresponding leads on the jumper chip, and then fixed using Epotek H70E.  Finally, electrical contact between the WH and the jumper chip was established using Epotek H20E with the aid of a hydraulic micromanipulator system (Narishige Three-axis Hanging Joystick Oil Hydraulic Micromanipulator, model: MMO-202ND) operated under an optical microscope. Finally, we used an Asylum Research Cypher MFM to confirm successful electrical control of the write pole magnetization (see Supplementary Figures 7-8).

\textit{Diamond nanowire tips:}
Single-crystal diamond nanowire (DNW) tips were fabricated via inductively-coupled plasma etching following procedures detailed in a previous publication \cite{tao15nl}.  The nanowires were transferred from their mother substrate to an intermediate silicon wafer chip via PDMS stamping (Gel-Pak 4 padding material).  Two tips were prepared for this study; tip A consisted of a diamond nanowire with an apex diameter of $\sim 40\unit{nm}$ coated by $15\unit{nm}$ of YF$_3$ ($\chi = -1.0\ee{-6}$ \cite{skuta14}) and $10\unit{nm}$ of Pt ($\chi = +2.6\ee{-4}$ \cite{hoare52}). These layers were deposited onto a silicon chip used for tip A via ebeam evaporation.  The YF$_3$ was deposited with future $^{19}$F NMR experiments in mind and had no role in this study, and the Pt was deposited to turn the nanowire paramagnetic as well as to screen electric charges.  Tip B was a bare diamond nanowire ($\chi = -2.2\ee{-5}$ \cite{heremans94}) with an apex diameter of $\sim 18\unit{nm}$.  High-resolution scanning electron micrographs of both tips are shown in Supplementary Figures 2-3.  Coated and bare DNWs were attached to silicon cantilevers using a micromanipulator system.  The custom cantilevers had a length of $120\unit{\um}$, a shaft width of $4\unit{\um}$ and a thickness of $120\unit{nm}$ \cite{chui03}.  The spring constant was $\kc=90\unit{\mu Nm^{-1}}$, the resonance frequencies were between $\fc=5-6\unit{kHz}$, and the quality factors were around $Q=30,000$ at $4\unit{K}$, resulting in a nominal force sensitivity of about $4\unit{aN}$ per root Hz.  Readers may wish to consult Ref. \cite{tao15nl} for fine details of the nanowire handling and transfer processes.

\textit{Sample positioning:}
Due to the extremely smooth, topographically featureless surface of the WH air-bearing surface, locating the write pole by the scanning cantilever was a challenge.  We tested a number of methods and found focused-ion beam (FIB) deposition of platinum dot markers to be the most reliable and efficient one.  Dots about $200$\,nm in diameter and $70$\,nm in height were written close to the write pole.  These dot features were easily identified when performing scans of the cantilever at constant height over the surface using the cantilever resonance frequency as image contrast (see Supplementary Figure 9). 

\textit{Force signal acquisition: }
All measurements discussed in the main text were performed in a cryostat operated at $4\unit{K}$ and in high vacuum ($<1\ee{6}\unit{mbar}$).
The WH driving current at $\fc/2$ was generated by an arbitrary waveform generator (NI PXIe-5451).  TTL pulses were used to synchronize the lock-in amplifier (Stanford Research SR830) to which the cantilever motional signal from the optical interferometer was sent.  A negative feedback loop was used to damp the cantilever to an effective operating $Q$ of about $300$ in order to shorten the sensor response time and to keep the driven cantilever oscillation below $\sim 1\unit{nm}$.  The force was calibrated by measuring the thermomechanical noise of the cantilever at $4\unit{K}$, and comparing the oscillation amplitude of the magnetically driven cantilever with the rms-amplitude of the thermal motion.  We have only calibrated the force for tip B, but not for tip A.  The phase of the force, which is available as a lock-in output, was calibrated by minimizing the imaginary part of the signal.

\textit{Finite element modeling: }
We carried out finite element simulations with COMSOL to validate the experimental results. The geometry of the write pole and the surrounding return shield were extracted from Fig. 1d (see Supplementary Figure 1 a and b). The return shield thickness was estimated to be $\sim 200$\,nm based on a focused ion beam cut into the write head surface. We assumed FeCo (with fixed magnetization $M_\mr{pole}$) as material for the write pole and NiFe ($\mu_r = 2 \cdot 10^3$) for the return shield \cite{tsang06,jayasekara98}. No parameters were adapted to fit the experimental results. The simulation uses a write pole magnetization of $\mu_0 M_\mr{pole}=1\unit{T}$ and the numerical values reported in Table 1 t were obtained by a scaling with the correct pole magnetization.  Supplementary Figure 1 c shows a profile cut through the write pole and return shield at $y=0$.


{\bf Acknowledgments:}
This work has been supported by the ERC through Starting Grant 309301, the DIADEMS program 611143 by the European Commission, by the Swiss NSF through the NCCR QSIT, and by the ETH Research Grant ETH-03 16-1. We thank B. Stipe for technical advice on the write head devices, P. Gambardella and M. S. Gabureac for access to the Asylum Research Cypher MFM, and U. Grob, C. Keck, B. Moores, H. Takahashi, J. Wrachtrup, O. Zilberberg and the ETH Physics Machine Shop for experimental help and frutiful discussions.

{\bf Author contributions:}
C.L.D. conceived the project.  A.E. and T.H. fabricated the devices, characterized them at room temperature, and performed finite element simulations.  Y.T. set up the experiment and carried out the measurements. C.L.D., Y.T., and A.E. analyzed the results and wrote the manuscript.




\newpage
{\noindent \bf Figure 1} \\

\newpage
{\noindent \bf Figure 2} \\

\newpage
{\noindent \bf Figure 3} \\

\newpage
{\noindent \bf Figure 4} \\

\newpage
{\noindent \bf Figure 5} \\

\newpage
{\noindent \bf Figure 6} \\

\newpage
{\noindent \bf Table 1} \\
\begin{table}[h!]
\centering
\begin{tabular}{p{2cm} p{2cm} p{2cm} p{2cm} p{2cm} p{2cm} p{2cm} }
\hline\hline
\       & \multicolumn{3}{l}{Magnetic field $|\vecB|$} & \multicolumn{3}{l}{Field gradient $\partial_x|\vecB|$} \\
\       & $12\unit{mA}$ & $30\unit{mA}$ & saturation & $12\unit{mA}$ & $30\unit{mA}$ & saturation \\
\hline
Experiment  & $0.51\unit{T}$  & $0.87\unit{T}$ & ---            & $17\unit{MTm}^{-1}$  & $28\unit{MTm}^{-1}$ & --- \\
Model       & $0.34\unit{T}$  & $0.64\unit{T}$ & 0.98$\unit{T}$ & $10\unit{MTm}^{-1}$  & $19\unit{MTm}^{-1}$ & $29\unit{MTm}^{-1}$\\
\hline\hline
\end{tabular}
\caption{\textbf{Pole field and gradient extracted from study.} Values were measured at $z=5\unit{nm}$ and are reported for drive currents of $12\unit{mA}$ (pole magnetization $0.83\pm0.07\unit{T}$), $30\unit{mA}$ ($1.57\pm0.08\unit{T}$), and for magnetic saturation ($2.4\unit{T}$), respectively.
The estimated error for the experimental values is $+28\%/-11\%$ due to the uncertainty in the tip volume.  The model values are probably slightly overestimated due to the few-nm thick protective layer of diamond-like carbon and Pt on top of the write pole.  See supplementary material for details.
}
\label{table:gradient}
\end{table}

\end{document}